\documentclass[aps,pra,reprint,superscriptaddress,showpacs,longbibliography,floatfix,footnoteinbib]{revtex4-1}
\usepackage{amsmath,amssymb,graphicx,units}
\usepackage[plainpages=false,pdfpagelabels,colorlinks=true,linkcolor=red,
urlcolor=blue,citecolor=blue,pdftitle={Title},pdfauthor={},pdfdisplaydoctitle=true,
pdfduplex=DuplexFlipLongEdge]{hyperref}

\newcommand{\ket}[1]{{\vert #1\rangle}}

\newcommand{\1}{\mbox{\bf 1}}

\makeatother
\begin{document}

\title{Interplay of local order and topology in the extended Haldane-Hubbard model}
\author{Can Shao}
\affiliation{Beijing Computational Science Research Center, Beijing 100084, China}

\author{Eduardo V. Castro}
\affiliation{Centro de F\'\i sica das Universidades do Minho e Porto, Departamento de F\'\i sica e Astronomia, Faculdade de Ci\^encias, Universidade do Porto, 4169-007 Porto, Portugal}

\affiliation{Beijing Computational Science Research Center, Beijing 100084, China}

\author{Shijie Hu}
\affiliation{Beijing Computational Science Research Center, Beijing 100084, China}

\author{Rubem Mondaini}
\email{rmondaini@csrc.ac.cn}
\affiliation{Beijing Computational Science Research Center, Beijing 100084, China}


\begin{abstract}
We investigate the ground-state phase diagram of the spinful extended Haldane-Hubbard model on the honeycomb lattice using an exact-diagonalization, mean-field variational approach, and further complement it with the infinite density matrix renormalization group, applied to an infinite honeycomb cylinder. This model, governed by both on-site and nearest-neighbor interactions, can result in two types of insulators with finite local order parameters, either with spin or charge ordering. Moreover, a third one, a topologically nontrivial insulator with nonlocal order, is also manifest. We test expectations of previous analyses in spinless versions asserting that once a local order parameter is formed, the topological characteristics of the ground state, associated with a finite Chern number, are no longer present, resulting in a topologically trivial wave function. Our study confirms this overall picture, and highlights how finite-size effects may result in misleading conclusions on the coexistence of finite local order parameters and nontrivial topology in this model.
\end{abstract}

\maketitle

\section{Introduction}\label{sec:intro}

Topological phases, which evade the paradigm of the conventional Landau-Ginzburg theory of spontaneously broken symmetries associated with the onset of a local order parameter, have been paramount to characterize and classify a large class of materials~\cite{Zhang19, Vergniory19, Tang19, Kruthoff2017}. In the past few years, the classification of topologically ordered states in noninteracting systems is believed to be complete~\cite{Schnyder08,Kitaev09}. Nonetheless, interacting topological models are expected to display much richer phenomena~\cite{Hohenadler13}, such as antiferromagnetic topological insulating states~\cite{Mong10, Fang13, Yoshida13, Miyakoshi13} or interaction-driven topological Mott insulators, in otherwise topologically trivial models~\cite{Raghu08, Wen10, Budich12, Dauphin12, Weeks10, LeiWang12, Ruegg11, Yang11, Yoshida14}. Some of these results, obtained via mean-field methods, have been disputed~\cite{Garcia_Martinez13, Daghofer14, Motruk15, Capponi15, Scherer15}, but two-dimensional systems with quadratic band crossings and weak interactions may yet allow the observation of interaction-induced nontrivial topology~\cite{Sun09, Murray14, Venderbos16, Wu16, Zhu16, Vafek10, Wang17}.

In the scope of strong interactions, topologically ordered states were seen to be absent when the system develops either charge or magnetic ordering \cite{Varney10, Varney11, Rachel10, Yamaji11, Zheng11, Yu11, Griset12, Hohenadler11, Hohenadler12, Reuther12, Laubach14}. Recently, however, a new class of exotic states has been shown, where in an interacting spinful version of the Haldane model it is possible to observe spontaneous SU(2) symmetry breaking. This is manifested as one spin species yet remaining topological, whereas the other turns trivial upon the increasing of a control parameter, resulting in a phase with Chern number equal to $1$~\cite{Vanhala16,Tupitsyn19,Mertz19,Wang19}.

In this paper, we further investigate the possibilities that finite local order parameters can coexist with a topological phase, obtaining the phase diagram of the half-filled spinful Haldane model (see Fig.~\ref{fig_1}), in the presence of both on-site and nearest-neighbor repulsive interactions on a honeycomb lattice. Sufficient local interactions are known to spontaneously induce a symmetry breaking~\cite{Meng10,Assaad13} resulting in an antiferromagnetic (Mott) insulator, whereas its nearest-neighbor counterpart induces a charge density wave (CDW) insulator if large enough, associated with a discrete (inversion) symmetry breaking. Our main finding is that, in general, when the development of either order occurs, the topological characteristics of the wave function, encoded on a finite Chern number, are no longer present. Exceptions to this, however, may occur in finite lattices, and only a careful investigation for large lattice sizes with suitable point group symmetries can clarify its occurrence in the thermodynamic limit \footnote{In passing, we notice that the pure extended Hubbard model on the honeycomb lattice has been studied by either cluster DMFT~\cite{Wu14} or via hybrid QMC techniques~\cite{Buividovich18}. In the latter, albeit sign problem-free, the observation of the CDW phase has been proven elusive, yet our method can clearly describe it.}.

One of the main challenges in investigating the interplay of topology and interactions is to unbiasedly compute the ground-state properties, and thus the topological invariants for the model of interest. If the model lacks time-reversal symmetry, as the Haldane-Hubbard model, quantum Monte Carlo (QMC) methods are largely limited due to the presence of a severe sign problem~\cite{Hirsch85,Loh90,Troyer05}. Cluster dynamical mean-field theory (DMFT), on the other hand, has been very successful~\cite{Wu16a,Vanhala16,Imriska16}, but the necessity to employ QMC as an impurity solver, also limits the low-temperature regime with larger cluster sizes, where again, a vanishing average sign is detrimental to simulations~\cite{Imriska16}.

Density matrix renormalization methods, on the other hand, are particularly reliable in investigating topological properties, but more easily applicable to ladder or cylinder geometries if beyond one dimension~\cite{Jiang12, Grushin15,Motruk15,Zhu15,Ren18,Barbarino19}. We have chosen thus a combination of three numerical methods. The first is the exact diagonalization (ED), which in spite of the small lattice sizes amenable to computations has been proven to be extremely useful in characterizing topological interacting systems~\cite{Varney10,Varney11,Jia13,Guo14,Wu16}. In particular, previous investigations indicate that on a honeycomb lattice, clusters with reciprocal lattices containing the $K$ high-symmetry point are able to grasp the fundamental critical features owing to the closing of the excitation gap at this point during the topological phase transition~\cite{Varney10,Varney11}. To put these results in perspective, we complement with a mean-field analysis of the model, qualitatively corroborating the onset of the ordering depending on the interaction parameters. Lastly, to finally identify whether or not a spontaneous SU(2) symmetry breaking can occur when approaching the thermodynamic limit (associated with a Chern number equal to $1$ as observed in \cite{Vanhala16,Tupitsyn19,Mertz19,Wang19}), we complement these results with the infinite density matrix renormalization group (iDMRG)~\cite{White92,Kjall13}.

The presentation is organized as follows: In Sec.~\ref{sec:model}, we introduce the model and all the quantities we use to characterize the different phases, using the three different numerical methods. Sections \ref{sec:ED_results} and \ref{sec:MF_results}, respectively, present the results using exact methods (ED and iDMRG) and mean field, respectively. Lastly, Sec.~\ref{sec:conclusion} summarizes and discusses the results.

\section{Model and measurements} \label{sec:model}
We study the extended Haldane-Hubbard model (EHHM), which is a combination of the Haldane model~\cite{Haldane88} and the extended Hubbard model on the honeycomb lattice~\cite{Wu14,Buividovich18},
\begin{eqnarray}
\hat H=&-&t_1\sum_{\langle i,j\rangle,\sigma}(\hat c^{\dagger}_{i,\sigma} \hat c^{\phantom{}}_{j,\sigma}+\text{H.c.}) \nonumber \\
&-&t_2\sum_{\langle\langle i,j\rangle\rangle,\sigma}(e^{{\rm i}\phi_{ij}}\hat c^{\dagger}_{i,\sigma} \hat c^{\phantom{}}_{j,\sigma}+\text{H.c.}) \nonumber \\
&+&U\sum_{i}\hat n_{i,\uparrow}\hat n_{i,\downarrow}+V\sum_{\langle i,j\rangle,\sigma,\sigma'}\hat n_{i,\sigma}\hat n_{j,\sigma'}.
\label{eq:H}
\end{eqnarray}
Here, $\hat c^{\dagger}_{i,\sigma}$ ($\hat c^{\phantom{}}_{i,\sigma}$) represents the electronic creation (annihilation) operator at site $i$ with spin $\sigma=\uparrow,\downarrow$, and $\hat n_{i,\sigma}\equiv \hat c_{i,\sigma}^\dagger\hat c_{i,\sigma}^{\phantom{}}$, the corresponding number operator. $t_1$ ($t_2$) denotes the nearest-neighbor (next-nearest-neighbor) hopping energy scale; $U$ and $V$ are the on-site and nearest-neighbor interactions, respectively.  A complex phase $\phi_{i,j}=\pm\phi$ representing the loops in the clockwise (anticlockwise) direction is added to the next-nearest-neighbor hopping term. This phase, if chosen such that $0<\phi<\pi$, originates a model which breaks time-reversal symmetry. In the non- and weak-interacting regimes, the ground state can thus be characterized by a topological invariant, the Chern number~\cite{Haldane88}. Throughout the paper, we focus on the ground-state phase diagram of Eq.~\eqref{eq:H}, at half filling, with periodic boundary conditions (PBCs), using both exact methods and mean-field techniques, in lattices containing $N$ unit cells (and $2N$ sites). In particular, when using iDMRG, we focus on an infinite-cylinder geometry, thus preserving PBCs along one direction while already assessing the thermodynamic limit on the other. Below, we briefly describe their methodology and how the observables are computed.

\subsection{Exact diagonalization in real space}
By employing periodic boundary conditions, we make use of translational symmetries, reducing the Hilbert space size by a factor of $N$. We proceed with a large-scale diagonalization, where we apply either Arnoldi~\cite{Lehoucq97arpack} or Krylov-Schur methods~\cite{Petsc,Slepc} to extract the ground state, and a few excited states of Eq.~\eqref{eq:H}, for lattices with up to $N=9$, i.e., 18 sites. As will later become clear, such lattice sizes are essential for the analysis, since clusters with a reciprocal lattice that contain the zone corner ($K$ high-symmetry point) could exhibit the characteristic first-order phase transition from the topological to the topologically trivial phase, while others may miss this feature, displaying it as a second-order one~\cite{Varney10,Imriska16}. In the next section, we report results for clusters encompassing both cases, and this will become more evident.

The characterization of the quantum phase transition is done via computing different quantities, such as the ground-state fidelity metric, the charge and spin structure factors, and the Chern number. The first is defined as~\cite{Zanardi06,CamposVenuti07,Zanardi07}
\begin{eqnarray}
g(x,\delta x)\equiv\frac{2}{N }\frac{1-|\langle \Psi_0(x)|\Psi_0(x+\delta x)\rangle|}{(\delta x)^2},
\label{eq:g}
\end{eqnarray}
where $x$ represents the interaction parameters $U$ or $V$, and $|\psi_0(x)\rangle$ [$|\psi_0(x+\delta x)\rangle$] the ground state of $\hat H(x)$ [$\hat H(x+\delta x)$]. This quantity is expected to produce a diverging peak with the system size, and has been routinely used to characterize different phase transitions, since it makes no underlying assumptions about the associated order parameter~\cite{Yang07,Varney11,Jia11,Mondaini15}. In what follows, we set $\delta x = 10^{-3}$ for either $x = U$ or $V$.

To probe the onset of the different local orders, with either spin-density wave (SDW) or charge-density wave (CDW), we define structure factors in a staggered fashion,
\begin{eqnarray}
S_{\text{SDW}} = \frac{1}{N}\sum\limits_{i,j}{(-1)}^{\eta}  \langle (\hat n_{i,\uparrow}-\hat n_{i,\downarrow}) (\hat n_{j,\uparrow}-\hat n_{j,\downarrow})\rangle, \nonumber \\
S_{\text{CDW}} = \frac{1}{N}\sum\limits_{i,j}{(-1)}^{\eta}  \langle (\hat n_{i,\uparrow}+\hat n_{i,\downarrow}) (\hat n_{j,\uparrow}+\hat n_{j,\downarrow})\rangle,
\label{eq:S}
\end{eqnarray}
with $\eta = 0$ ($\eta = 1$) if sites $i$ and $j$ are in the same (different) sublattice, i.e., A or B.

The topological invariant is quantified by the Chern number. If using twisted boundary conditions (TBCs)~\cite{Didier91}, it can be defined as an integration over the Brillouin zone~\cite{Niu85},
\begin{align}
  C = \int \frac{d\phi_x d\phi_y}{2 \pi {\rm i}} \left( \langle\partial_{\phi_x}
      \Psi^\ast | \partial_{\phi_y} \Psi\rangle - \langle{\partial_{\phi_y}
      \Psi^\ast | \partial_{\phi_x} \Psi\rangle} \right),
\label{eq:C}
\end{align}
with $\ket{\Psi}$ being the many-particle wave function, and $\phi_x$ ($\phi_y$) the  twisted phase along the $x$ ($y$) direction. Provided there are no degeneracies in the ground-state manifold $E_0(\phi_x,\phi_y)$, Eq.~\eqref{eq:C} results in a $\mathbb{Z}$ integer number. An immediate drawback is that this expression requires the computation of derivatives and integrals of the wave function with respect to the continuous variable. It has been shown, however, to already converge to the true Chern number if using a sufficiently discretized version~\cite{Fukui05, Varney11, Zhang13}. In what follows, we report results using a mesh of $6\times 6$ phases $(\phi_x,\phi_y)$ over the Brillouin zone. A comparison with finer meshes is exemplified in the Appendix.

\subsection{Infinite density matrix renormalization group} \label{sec:iDMRG}
We further check the quantized Hall conductance in some regions of the phase diagram for an infinitely long cylinder at zero temperature, with the goal of characterizing the Chern number $C$ of the ground-state wave function via a charge pumping scheme. This can be realized by the infinite-DMRG method in a natural way when employing a geometry with finite circumference $L_y$ and an infinite $L_x$. After inserting a finite $\theta$ flux along the positive/negative $x$ axis, a certain number of charges are pumped to the left/right side and the statistic of the accumulative discrepancy is defined as~\cite{Grushin15}
\begin{equation}
Q(\theta) = \left|\sum_{l} \Lambda^2_l(\theta) (Q^{L}_{l} (\theta) - Q^{R}_{l} (\theta))\right|,
\end{equation}
where $\Lambda_{l}$ is the singular value after decomposition of the whole cylinder into two semi-infinite parts, $Q^{L/R}_{l}$ is the charge degree of freedom on the left/right side marked for the $l$-th renormalized basis, and $l$ runs over all truncated bond dimension $m$. In general, $Q(0)\ne 0$ if the ground state breaks the inversion symmetry, such as in the CDW phase of the model we study here.

\subsection{Mean-field method in momentum space}
To contrast the results obtained via ED and iDMRG, we report in Sec.~\ref{sec:MF_results} mean-field (MF) calculations. For that, we employ a two-site unit cell computing the corresponding fields in momentum space, owing to the translational invariance of the problem. We choose $\textbf{a}_1 = a(-\frac{1}{2},\frac{\sqrt{3}}{2})$ and $\textbf{a}_2 = a(\frac{1}{2},\frac{\sqrt{3}}{2})$ as the basis vectors in real space; their counterparts in reciprocal space are $\textbf{b}_1 = \frac{1}{a}(-2\pi,\frac{2\pi}{\sqrt{3}})$ and $\textbf{b}_2 = \frac{1}{a}(2\pi,\frac{2\pi}{\sqrt{3}})$. By introducing the operators $a_{\textbf{k},\sigma}^{\dag}=\frac{1}{\sqrt{N}}\sum_{i\in \text{A}}c_{i,\sigma}^{\dag}e^{\mathrm{i}\textbf{k}\cdot\textbf{r}_i}$ and $b_{\textbf{k},\sigma}^{\dag}=\frac{1}{\sqrt{N}}\sum_{i\in \text{B}}c_{i,\sigma}^{\dag}e^{\mathrm{i}\textbf{k}\cdot\textbf{r}_i}$, the Hamiltonian \eqref{eq:H} can be expressed as follows:
\begin{align}
\hat H = \hat H_{\text{0}} + \hat H_{\text{I}},
\label{eq:Hk}
\end{align}
with,
\begin{align}
\hat H_{\text{0}}=\sum_{\textbf{k},\sigma} \left( m_+(\textbf{k})a_{\textbf{k},\sigma}^{\dag}a_{\textbf{k},\sigma} + m_-(\textbf{k})b_{\textbf{k},\sigma}^{\dag}b_{\textbf{k},\sigma}
\notag\right.
\\
\phantom{=\;\;}
\left.-t_1 g(\textbf{k})a_{\textbf{k},\sigma}^{\dag}b_{\textbf{k},\sigma}-t_1 g^*(\textbf{k})b_{\textbf{k},\sigma}^{\dag}a_{\textbf{k},\sigma}\right),
\label{eq:H0}
\end{align}
and,
\begin{align}
\hat H_{\text{I}}&=\frac{U}{N}\sum_{\textbf{k},\textbf{k'},\textbf{q}} c_{\textbf{k+q},\uparrow}^{\dag}c_{\textbf{k},\uparrow}c_{\textbf{k}'-\textbf{q},\downarrow}^{\dag}c_{\textbf{k}',\downarrow} \nonumber \\
&+\frac{V}{N}\sum_{\sigma,\sigma'}\sum_{\textbf{k},\textbf{k'},\textbf{q}} g(\textbf{q}) a_{\textbf{k+q},\sigma}^{\dag}a_{\textbf{k},\sigma}b_{\textbf{k}'-\textbf{q},\sigma'}^{\dag}b_{\textbf{k}',\sigma'},
\label{eq:HI}
\end{align}
where $g(\textbf{k})=1+e^{-\mathrm{i}\textbf{k}\cdot \textbf{a}_1}+e^{-\mathrm{i}\textbf{k}\cdot \textbf{a}_2}$, and $m_{\pm}(\mathbf{k})= -2t_2\left[\cos({\bf k}\cdot {\bf a}_1\mp\phi) \allowbreak +\cos({\bf k}\cdot {\bf a}_2\pm\phi) \allowbreak + \allowbreak \cos({\bf k}\cdot ({\bf a}_1-{\bf a}_2)\pm\phi)\right]$.

After a mean-field decoupling of the four-fermion terms (including both Hartree and Fock terms), we arrive at the following mean-field Hamiltonian,
\begin{equation}
\begin{split}
\hat H_{\rm MF}=&\hat H_0 \nonumber\\
+&\sum_{\textbf{k}}\psi_{\textbf{k}}^{\dag} \left(
  \begin{array}{cccc}
    \varepsilon_{\uparrow}^{a} & \xi_{\uparrow\uparrow}(\textbf{k}) & \varepsilon_{\uparrow\downarrow}^{a}& \xi_{\uparrow\downarrow}(\textbf{k})\\
    \xi^*_{\uparrow\uparrow}(\textbf{k}) & \varepsilon_{\uparrow}^{b} & \xi^*_{\downarrow\uparrow}(\textbf{k}) &\varepsilon_{\uparrow\downarrow}^{b}\\
    (\varepsilon_{\uparrow\downarrow}^{a})^* & \xi_{\downarrow\uparrow}(\textbf{k}) & \varepsilon_{\downarrow}^{a} & \xi_{\downarrow\downarrow}(\textbf{k})\\
    \xi^*_{\uparrow\downarrow}(\textbf{k}) & (\varepsilon_{\uparrow\downarrow}^{b})^* & \xi^*_{\downarrow\downarrow}(\textbf{k}) & \varepsilon_{\downarrow}^{b}
  \end{array}
\right)\psi_{\textbf{k}}
\end{split}
\end{equation}
where we have used the spinor notation $\psi^\dagger_{\textbf{k}}=[a^\dagger_{\textbf{k},\uparrow},b^\dagger_{\textbf{k},\uparrow},a^\dagger_{\textbf{k},\downarrow},b^\dagger_{\textbf{k},\downarrow}]$ as a basis for each lattice momentum ${\bf k}$. Now, by making use of the variational mean-field approach, we end up with the following set of mean-field equations, which complemented by the charge conservation, need to be solved self-consistently:
\begin{align}
\xi_{\sigma\sigma'}(\textbf{k})&=-\frac{V}{N}\sum_{\textbf{q}}g(\textbf{k}-\textbf{q})\langle b_{\textbf{q},\sigma'}^{\dag} a_{\textbf{q},\sigma}\rangle_{\rm MF},\nonumber \\
\varepsilon_{\sigma}^{a}&=Un_{-\sigma}^{a}+3V\sum_{\sigma'} n_{\sigma'}^{b},\nonumber \\
\varepsilon_{\sigma}^{b}&=Un_{-\sigma}^{b}+3V\sum_{\sigma'} n_{\sigma'}^{a},\nonumber \\
\varepsilon_{\uparrow\downarrow}^{a}&=-\frac{U}{N}\sum_{\textbf{q}}\langle a_{\textbf{q},\downarrow}^{\dag} a_{\textbf{q},\uparrow}\rangle_{\rm MF},\nonumber \\
\varepsilon_{\uparrow\downarrow}^{b}&=-\frac{U}{N}\sum_{\textbf{q}}\langle b_{\textbf{q},\downarrow}^{\dag} b_{\textbf{q},\uparrow}\rangle_{\rm MF},
\end{align}
with densities $n_{\sigma}^{a}=\frac{1}{N}\sum_{\textbf{q}}\langle a_{\textbf{q},\sigma}^{\dag} a_{\textbf{q},\sigma}\rangle_{\rm MF}$ and $n_{\sigma}^{b}=\frac{1}{N}\sum_{\textbf{q}}\langle b_{\textbf{q},\sigma}^{\dag} b_{\textbf{q},\sigma}\rangle_{\rm MF}$. In the expressions above, the averages $\langle \cdots \rangle_{\rm MF}$ are taken in the grand-canonical ensemble by accounting for the Boltzmann factor in the mean-field Hamiltonian.

Once convergence for the free energy has been achieved, we can compute the CDW and SDW order parameters,
\begin{eqnarray}
\mathcal{O}_{\text{CDW}} &=& \left|\left( n^{a}_{\uparrow}+ n^{a}_{\downarrow})-( n^{b}_{\uparrow}+ n^{b}_{\downarrow}\right)\right|,\nonumber \\
\mathcal{O}_{\text{SDW}} &=& \left|\frac{1}{2}\left(\langle \vec  S_a \rangle_{\rm MF} - \langle \vec S_b\rangle_{\rm MF} \right)\right|,
\label{eq:O}
\end{eqnarray}
where $\vec S_i = \frac{1}{2}\sum_{\alpha\beta}c_{i\alpha}^\dagger \vec\sigma_{\alpha\beta}c^{\phantom{\dagger}}_{i\beta}$, and $\vec \sigma = (\sigma^x, \sigma^y, \sigma^z)$ is the vector of spin-1/2 Pauli matrices. Even though we compute other order parameters related to different broken-symmetry phases, these two turned out to be the most stable. For the calculation of the Chern number, we use the discrete formulation in its multiband (non-Abelian) version~\cite{Fukui05}. In what follows, $t_1$ is set to be the unit of energy and $t_2=0.2$. We further fix the Haldane phase $\phi$ to $\pi/2$, in order to maximize the Chern insulating (CI) phase~\cite{Haldane88, Varney10}.

\section{Results of the exact methods}
\label{sec:ED_results}

\begin{figure}[t]
\centering
\includegraphics[width=0.5\textwidth]{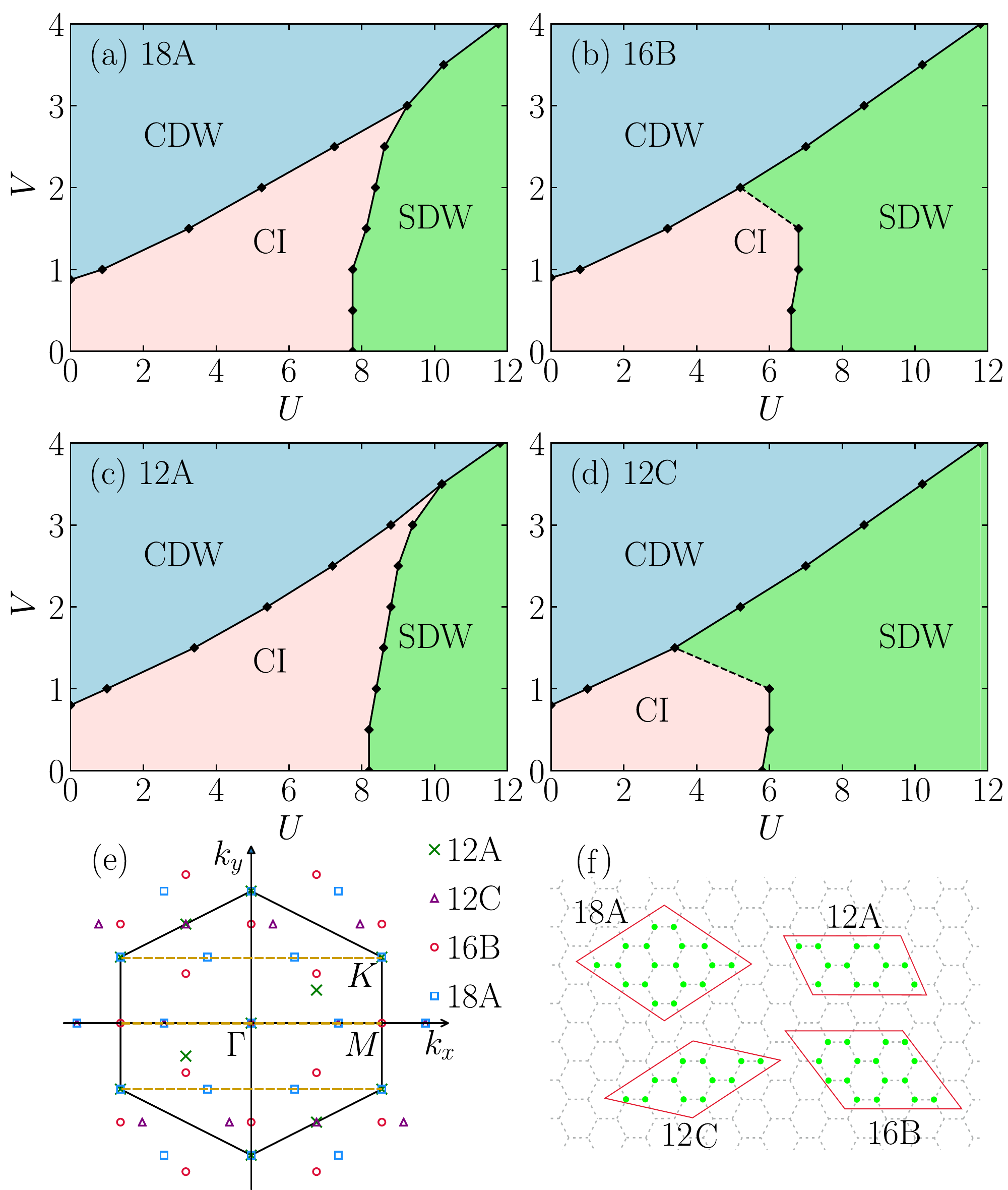}
\caption{[(a)-(d)] Phase diagrams in the parametric space ($U$, $V$) of the extended Haldane-Hubbard model on the 18-site (a) and 16-site (b) lattice, and two types of 12-site lattices (c) and (d), respectively, based on the ED calculations. As with elsewhere in the paper, the parameters are $t_2=0.2$ and $\phi=\pi/2$. The set of $k$ points for each finite cluster is schematically represented in (e) along with the marked high-symmetry points $\Gamma$, $M$, and $K$. The horizontal dashed lines depict the continuous set of momentum values attainable in the infinite-cylinder geometry  with $L = 6$. Panel (f) depicts the clusters used in the ED analysis; all such clusters form a bipartite lattice in the presence of PBCs, but only clusters 12A and 18A contain the $K$ Brillouin zone corner as a valid momentum point.}
\label{fig_1}
\end{figure}

We start by directly presenting the phase diagram [Fig.~(\ref{fig_1})] obtained via ED using four different clusters $18$A, $16$B, $12$A and $12$C [see Fig.~\ref{fig_1}(f)], and the particular characterization of each phase will be presented afterward. These clusters are selected in such a way that they are able to accommodate a N\'eel state (i.e., they are bipartite if considering PBCs), and we notice that clusters 12A and 18A also exhibit the $K$ point as a valid momentum point in the reciprocal lattice, unlike clusters 12C and 16B [Fig.~\ref{fig_1}(e)]. These can result in systematic finite-size effects, as we argue below.

The phase diagrams are characterized by phases with the formation of a local order parameter as a result of the interactions: large $U$ and $V$ result in SDW and CDW phases, respectively. At the regime of weak interactions, a topologically nontrivial CI phase with Chern number $C=2$ distributes at a closed area. That is, the parent spinful noninteracting model possess a ground state characterized by a topological invariant which survives in the presence of both interactions, as long as they are sufficiently small or compete, preventing the onset of the formation of a local order parameter. The transition to a charge-ordered phase is fairly consistent across the different system sizes, whereas the one to a spin-ordered one suffers from slightly larger finite-size effects. These finite-size effects are not merely related to the number of sites in the lattice but rather if the cluster under study contains or not the $K$ point in its reciprocal lattice. For that reason, clusters 12A and 18A present a quantitatively similar SDW transition, while for 12C and 16B this transition is systemattically deviated to smaller on-site interactions $U$.

\begin{figure}[t]
\centering
\includegraphics[width=0.5\textwidth]{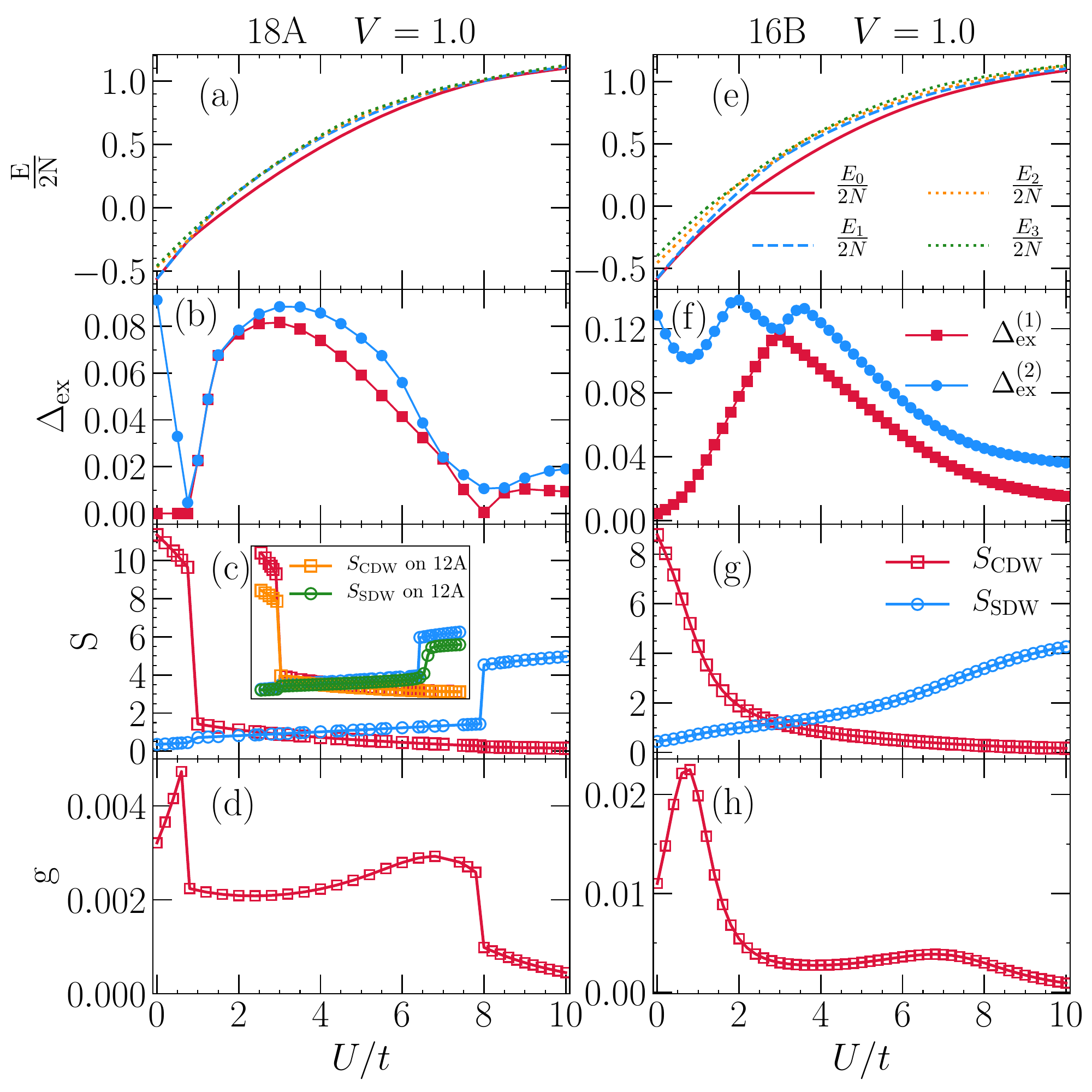}
\caption{Four lowest-lying energy levels $E_{\alpha}$ [(a) and (e)], excitation gaps $\Delta E^{(\alpha)_{\rm ex}}$ [(b) and (f)], structure factors $S$ [(c) and (g)], and the fidelity metric $g$ [(d) and (h)] of the extended Haldane-Hubbard model with $V=1$ on the 18-site (left panels) and 16-site (right panels) clusters. First-order phase transitions are only seen for the 18A cluster (see text). The inset in panel (c) includes the structure factors for the cluster 12A, highlighting the extensive nature of the corresponding correlators' sum within the ordered phases.}
\label{fig_2}
\end{figure}

These phase diagrams are constructed from the analysis of several quantities mentioned in Sec. \ref{sec:model}. In order to display all relevant features, we focus on a typical line with $V=1$, which successively crosses the CDW, CI, and SDW phases when increasing $U$ from $U=0$ to $U=10$. The results are shown in Fig. \ref{fig_2} for cluster 18A on the left panels, whereas the right panels display the corresponding quantities for cluster 16B. The lattices with 12 sites present similar results to their larger counterparts, in what concerns the presence or absence of the previously mentioned $K$ point.

To start, we characterize the type of the transition, by analyzing the low-lying energy spectrum dependence across the different phases. The first four energy levels, i.e., $E_{\alpha}$ with $\alpha=0,1,2,3$ ($E_0$ is the ground-state energy), are plotted in Figs. \ref{fig_2} (a) and \ref{fig_2} (e). A careful inspection shows that level crossings occur at the two phase boundaries (CDW-CI and CI-SDW) for the 18A cluster, resulting in first-order phase transitions. These are absent in the 16B cluster, and we reemphasize the carefulness required in selecting lattices with the most suitable point-group symmetries. These crossings are more easily identified if defining the excitation gaps $\Delta_{\text{ex}}^{(\alpha)}=(E_\alpha-E_0)/L$, as shown in Figs.~\ref{fig_2} (b) and \ref{fig_2} (f) for $\alpha = 1$ and 2, and clusters 18A and 16B, respectively. In the CI-SDW transition for the 18A cluster, due to the fact that the ground state is nondegenerate, a vanishing $\Delta^{(1)}_{\rm ex}$ precisely marks the phase boundary.

The CDW-CI transition, on the other hand, is more easily characterized by the vanishing of the second excitation gap, $\Delta^{(2)}_{\rm ex}$. The reason behind this is that in the CDW phase, the ground state is twofold degenerate in the thermodynamic limit (and nearly degenerate in the finite cluster we deal with), with a level crossing occurring between $E_2$ and $E_0$ (or $E_1$) as the transition is approached. In contrast, in the 16B cluster, such many-body gaps never close, but the transitions can be yet pinpointed by peaks in the fidelity metric $g$, displayed in Fig.~\ref{fig_2} (h). In turn, for the 18A cluster [Fig.~\ref{fig_2} (d)] a proper peak is missing (a discontinuity is instead observed) precisely due to the fact the transition is first order, and one needs a resolution of the control parameter (in this case $U$) that is small enough to capture the very narrow $\delta U$-dependent peak, $g_{\rm peak} = 2/\left(N \delta U^2\right)$~\cite{Varney11}. Again for the 16A cluster, the `hump' depicting the CI-SDW transition becomes wider and smaller in magnitude for larger $V$, making the characterization of this transition more challenging [see dashed lines in Figs.~\ref{fig_1}(b) and \ref{fig_1}(d)], and thus accounting for the difference between the phase diagrams of clusters containing or not the $K$ point.

Lastly, we report in Figs.~\ref{fig_2} (c) and \ref{fig_2} (g) the structure factors corresponding to each order, CDW and SDW, which display a characteristic discontinuous behavior as similarly found elsewhere for other models manifesting a transition between topologically nontrivial and trivial phases \cite{Varney10,Varney11,Wu16}. The inset in Fig.~\ref{fig_2} (c) shows the extensive nature of $S_{\rm CDW}$ and $S_{\rm SDW}$ within each phase, by contrasting the 18A cluster with its counterpart that also contains the $K$ point in its reciprocal space, the lattice 12A.

\begin{figure}[t]
\centering
\includegraphics[width=0.5\textwidth]{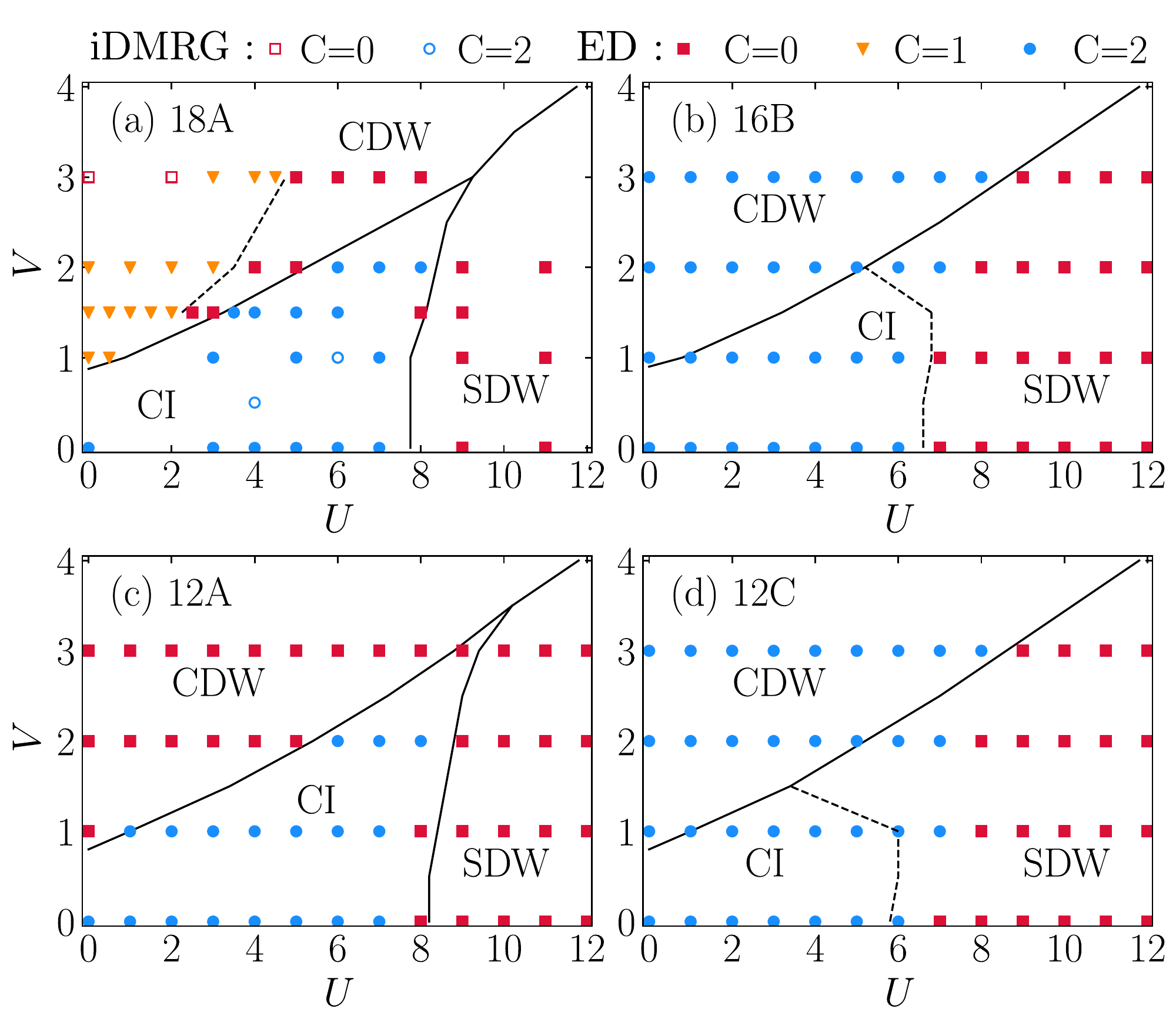}
\caption{Chern number results overlaid with the phase boundaries originally displayed in Fig.~\ref{fig_1} for the 18A (a), 16B (b), 12A (c), 12C (d) lattices. The solid square, triangle, and circle markers represent $C=0$, $1$, and $2$, respectively, obtained via ED. The corresponding empty markers in panel (a) denote the iDMRG results obtained on an infinite-cylinder geometry with $L=6$. The dashed line in panel (a) is drawn as a guide to the eye, delimiting the region where the $C=1$ regime is found within the CDW phase using ED, and not confirmed by iDMRG.
}
\label{fig_3}
\end{figure}

Back to the phase diagram in Fig.~\ref{fig_1}, we are now in position to characterize the phase that displays a nonlocal order parameter, the topologically nontrivial Chern insulating phase. We present in Fig.~\ref{fig_3} an overlay of the computed Chern number, using a discrete version of Eq.\eqref{eq:C} (See also the Appendix), and the original boundaries for the phases presented in Fig.~\ref{fig_1}. For the clusters 12C and 16B, a $C = 2$ phase gives way to a $C=0$ (thus topologically trivial) roughly at the same positions as the fidelity peak signals the CI-SDW phase transition, at large $U$ values. For the CDW phase, on the other hand, such change of the topologically invariant number is not seen in the ranges of $V$'s computed, but in the spinless version of the present model, such coexistence of a $C\neq0$ with a local order parameter has been attributed to the fact that $K$ is not in the set of $k$ points available for some finite clusters, precisely as here~\cite{Varney10,Varney11}.

The most interesting features of this computation are thus the ones that come from the calculation in clusters 12A and 18A, which contain the $K$ in its reciprocal lattice. Although for the 12A case [Fig.~\ref{fig_3}(c)] the computed Chern number closely follows the general belief that once the local order parameters develop the topological characteristics vanish, results from cluster 18A [Fig.~\ref{fig_3}(a)] are much richer.
In this case, we find that in part of the CDW phase, at smaller $U$ values, there is a coexistence regime of a finite CDW order with the presence of a SU(2) symmetry-breaking $C=1$ phase, observed in variants of this model that include a staggered on-site energy term~\cite{Vanhala16,Tupitsyn19,Mertz19,Wang19}. A physical interpretation of this result is that in such area of the phase diagram one pseudospin component of the spinful model remains topologically nontrivial whereas the other does not.

Whether this coexistence also occurs in the thermodynamic limit cannot be clarified by the ED method, specially in view of the fact that the cluster 18A does not possess all point group symmetries of the honeycomb lattice, albeit clearly possessing one of the essential ingredients, the manifestation of the $K$-point physics. In the Appendix, we further present more details on the numerical computation of the Chern number using the ED, in particular for this lattice size, highlighting that the presence of the $C=1$ phase is neither due to a potentially coarse discretization of the Berry curvature in Eq.\eqref{eq:C} nor that the first-excitation gap $\Delta E^{(1)}_{\rm ex}$  of the doublet states in this phase closes when performing the summation, which could eventually explain the unexpected results displayed in Fig.~\ref{fig_3}(a).

\begin{figure}[t]
\centering
\includegraphics[width=0.46\textwidth]{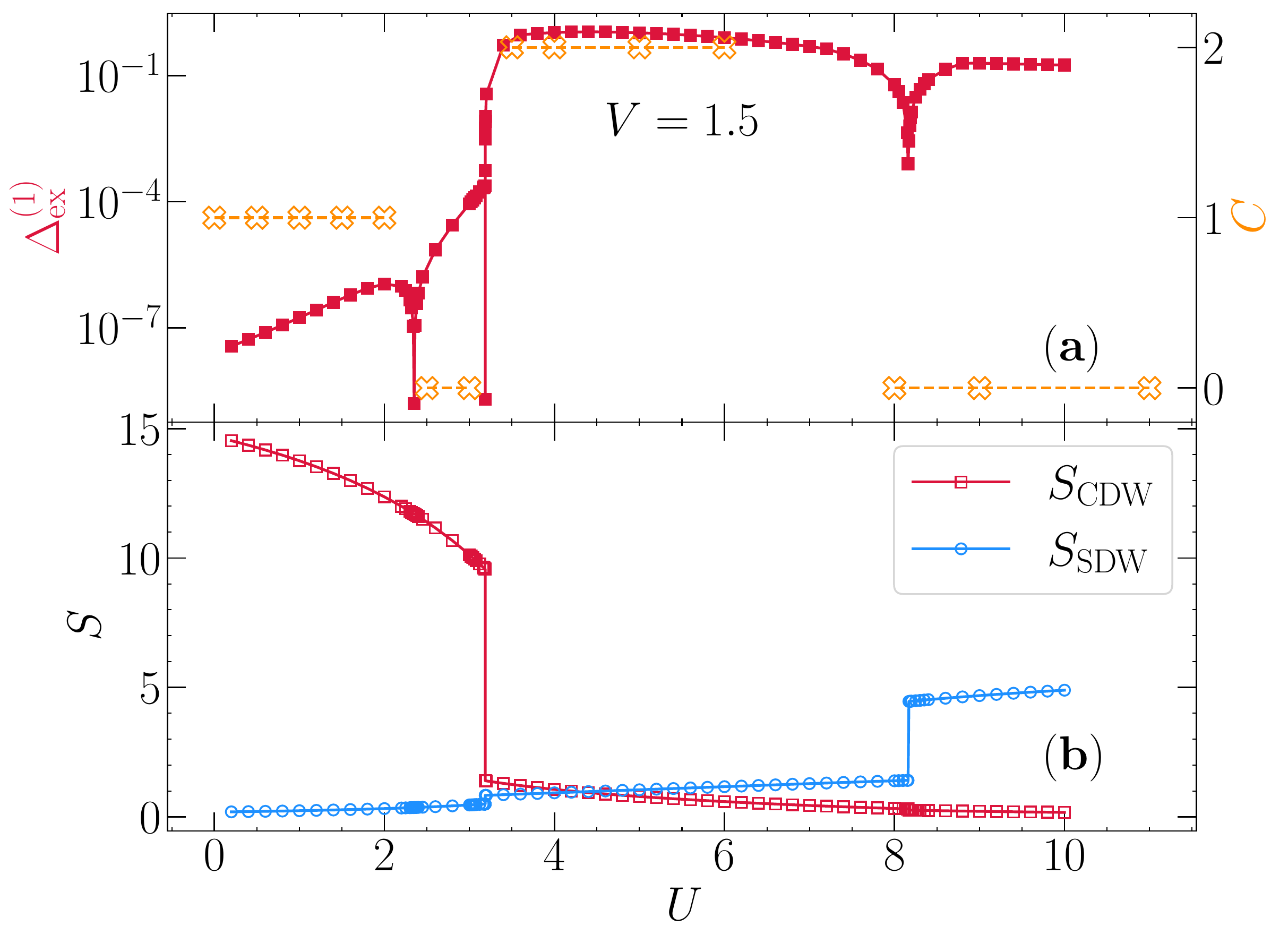}
\caption{(a) First-excitation gap $\Delta_{\rm ex}^{(1)}$ and Chern number along the $V=1.5$ cut. In (b), the corresponding ground-state structure factors, to facilitate the visualization of the transitions.
}
\label{fig_3_5}
\end{figure}

Yet, the $C=1$ to $C=0$ transition within the CDW phase for this cluster size requests further inspection. Figure~\ref{fig_3_5}(a) shows $\Delta_{\rm ex}^{(1)}$ along the $V=1.5$ line. There, the Chern number changes its value precisely because the doublet states display a level crossing, before the CDW-CI takes place with a second, first-order phase transition.

\begin{figure}[b]
\centering
\includegraphics[width=0.45\textwidth]{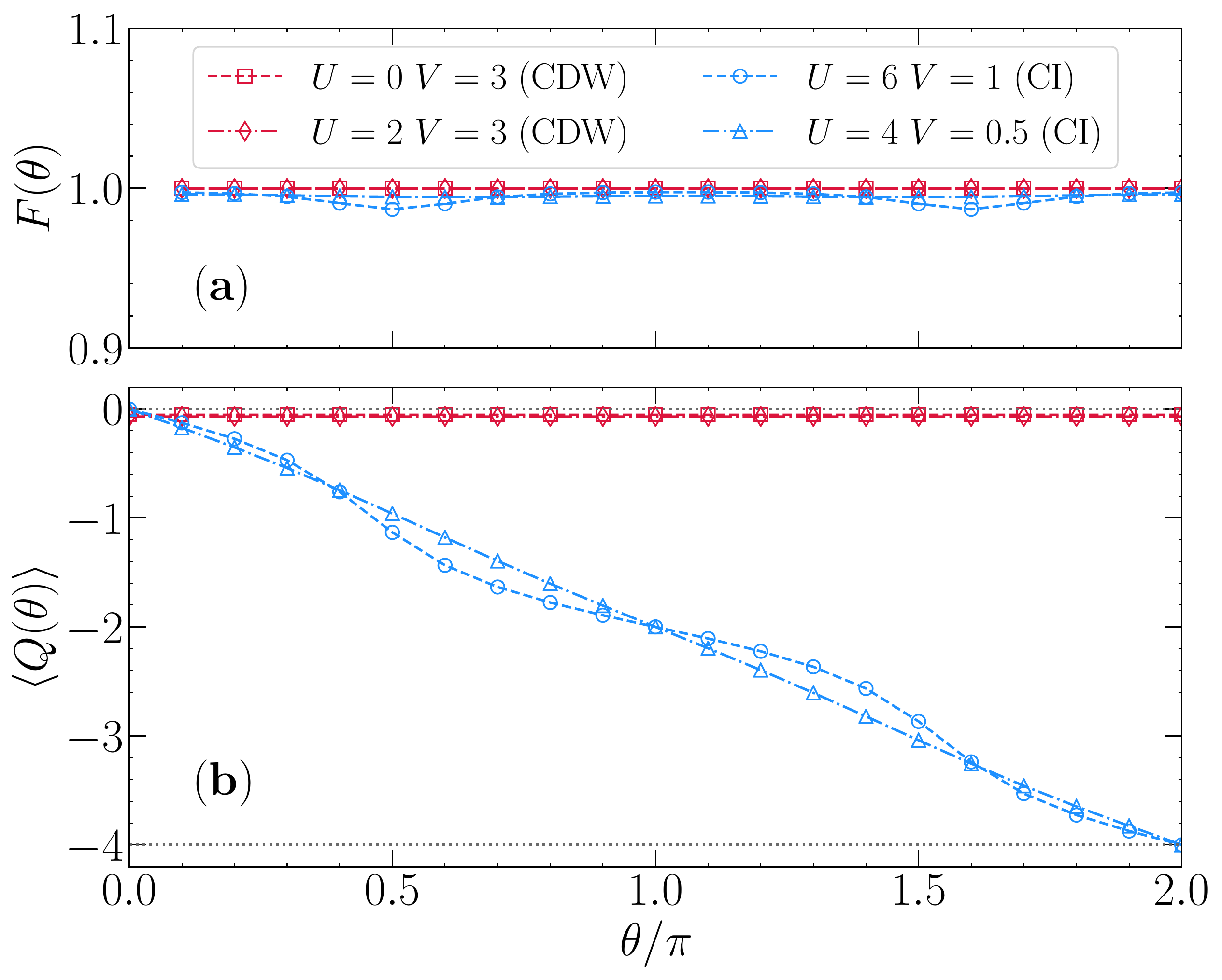}
\caption{(a) Fidelity $F$ and (b) accumulative discrepancy $Q$ as a function of the inserted magnetic flux $\theta$ in an adiabatical way with a small interval $\delta \theta/\pi=0.1$. Here $L_y=3$ (or $L=6$) and the truncated bond dimension $m=1024$.
}
\label{fig_3b}
\end{figure}

Nonetheless, by using the scheme of pumped charges (see Sec.~\ref{sec:iDMRG}) on an infinite cylinder, composed of three unit cells along its circumference ($L_y=3$), we finally clarify that the apparent SU(2)-symmetry-broken phase with $C=1$ is likely due to a finite-size effect. To monitor the adiabatical insertion of a magnetic flux, we define an overlap between two wave functions at $\theta$ and $\theta+\delta\theta$, the fidelity $F(\theta) = |\langle \Psi_0(\theta)|\Psi_0(\theta+\delta\theta)\rangle|$. Here the interval $\delta \theta$ is set at a small value, such as $\delta \theta/\pi=0.1$ in Fig.~\ref{fig_3b}, in order to preserve adiabaticity along the process, which has $F\approx 1$ [Fig.~\ref{fig_3b}(a)]. The pumping procedure results in different outcomes depending on topological characteristics of the underlying phase. For example, the adiabatic insertion of a $2\pi$ flux in the CDW phase does not pump any charges, i.e., $\Delta Q=Q(2\pi)-Q(0)=0$, yielding $C=0$, thus topologically trivial if $V=3$ and $U=0$ or 2 [Fig.~\ref{fig_3b}(b)]. On the hand, in the CI phase, the pumping charge $\Delta Q=4$ confirming $C=2$. As a result, no $C=1$ has been found for the parameters investigated using this scheme.

\section{Results of the mean-field calculation}\label{sec:MF_results}

\begin{figure}[t]
\centering
\includegraphics[width=0.5\textwidth]{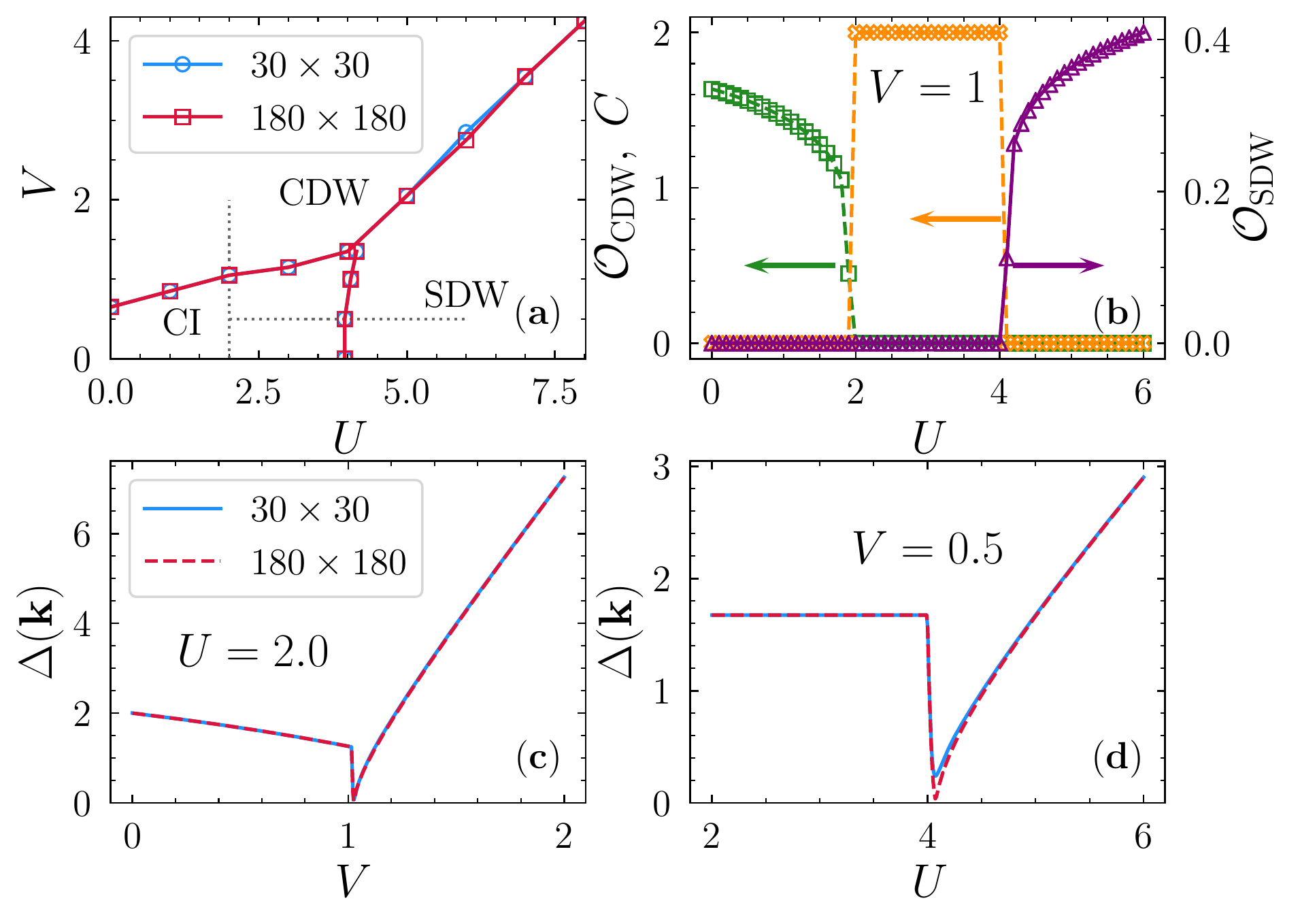}
\caption{(a) The phase diagram of Eq.~\ref{eq:H} under the mean-field approximation. (b) A line cut of the phase diagram at $V = 1$, showing the local order parameters ${\cal O}_{\rm CDW}$ and ${\cal O}_{\rm SDW}$ together with the corresponding Chern number $C$. Panels (c) and (d) display the smallest band gap $\Delta({\bf k})$ along the lines $U = 2$ and $V = 0.5$, respectively, and are  highlighted by the dotted lines in panel (a).}
\label{fig_4}
\end{figure}

To contrast the previous results, and to further understand finite-size discrepancies in the ED results for the Chern number in comparison to iDMRG ones, we now report the outcomes when casting the interactions in a mean-field form (see Sec.~\ref{sec:model}). The phase diagram, constructed by taking into account the onset of the order parameters ${\cal O}_{\rm CDW, SDW}$ [Eq.\eqref{eq:O}], is shown in Fig.~\ref{fig_4} (a) on $180\times180$ and $30\times30$ lattices. We first notice that the finite-size effects are rather small, and the phase boundaries are qualitatively very similar to the ones obtained from the ED method; i.e., the Chern insulating phase gives way to SDW or CDW once the on-site or nearest-neighbor interactions are sufficiently large.

A line cut in this phase diagram with $V = 1$ clearly demonstrates the different phases and associated orders, and is shown in Fig.~\ref{fig_4}(b). The sharpness of the transitions indicates that the coexistence of a topological phase with the formation of a local order parameter is reduced to a very small region close to the transition line, unlike presented in some of the previously shown ED results.  To see how this is connected with the general picture of the change of a topological invariant after a single-particle gap closing when increasing a control parameter, we show in Figs.~\ref{fig_4}(c) and \ref{fig_4}(d) the gap $\Delta({\bf k}) = \min[E_2({\bf k^\prime})-E_1({\bf k^\prime})]$ along line cuts $U = 2$ and $V = 0.5$, respectively. 
We first observe that the gap is not very sensitive to changes either in $V$ or $U$ inside the CI phase. The sharp drop in the gap at a critical value of the control parameter signals the onset of the local order parameters. After that, the gap closes quickly (with small finite-size corrections) and it does corroborate the change of the topological invariant at these points shown in Fig.~\ref{fig_4}(b). It is in the reduced region where the gap sharply drops to zero that long-range order and nontrivial topology coexist. Note, however, that despite its sharpness, the gap closes continuously and the phase transition into ordered phases is seemingly second order at the MF level. Nonetheless, this gap closing occurs, in both transitions, around the Dirac point, ${\bf k} \simeq K$.

\section{Summary and Discussion}\label{sec:conclusion}

We studied the extended Haldane-Hubbard model in the honeycomb lattice, at unity filling. Depending on the magnitude of the repulsive interactions (either on-site or nearest-neighbor) the ground state displays insulating behavior, with the presence of phases with finite local order parameters, as charge-density and (antiferromagnetic) spin-density waves, in addition of a topological phase, with its corresponding non local order parameter associated with a topological invariant. Besides, the transitions among all such different phases are first order, when computed with ED~\footnote{In mean-field, however, the continuous closing of the gap (on a small scale of the control parameter) is suggestive of a second order transition.}. This picture is numerically inferred in small lattices employing the exact diagonalization, and complemented by a variational mean-field analysis. In the former method, due to the small system sizes amenable to computations, some of the clusters may not display all the point group symmetries of the lattice in the thermodynamic limit. For that reason, we must caution that some of the results we present might suffer from systematic finite-size effects. Among those, a surprising result is the manifestation of an SU(2) symmetry breaking, where the appearance of a $C=1$ phase concurs with a charge-density wave in a large part of the phase diagram, when dealing with the largest cluster manageable to ED calculations. Such phase is absent in the variational mean-field results, and we show to be also \textit{absent} when computing the topological charge in infinite cylinders, using iDMRG: a demonstration of how dramatic finite-size effects can alter conclusions obtained in finite clusters.

This symmetry-broken phase, however, has been described in the Haldane-Hubbard model (i.e., with finite $U$ and $V=0$) in the presence of a staggered potential $\Delta$, sandwiched in between the standard antiferromagnetic Mott-insulating phase at large on-site interactions, and a band-insulating one~\cite{Vanhala16,Tupitsyn19}. An immediate investigation one could follow would be to test these predictions using infinite cylinders, and we leave this for future studies. In particular, a topological model in the presence of both $U$ and $V$ interactions such as ours has anticipated that by using mean-field calculations, the $C=1$ phase is yet manifest for a finite staggered potential~\cite{Wang19}.

\begin{acknowledgments}
The authors acknowledge insightful discussions with P.~Sacramento, H.~Guo, H.~Lu, and H.-Q.~Lin. C.S. acknowledges support from the China Postdoctoral Science Foundation (Grant No. 2019M650464). R.M. acknowledges support from NSFC Grants No. 11674021, No. 11851110757, and No. 11974039. C.S., S.H., and R.M. further acknowledge support from Grant NSAF-U1930402. The computations were performed on the Tianhe-2JK at the Beijing Computational Science Research Center (CSRC).
\end{acknowledgments}

\appendix

\section{The $C=1$ phase for cluster 18A} \label{sec:app_c_1}
We argue in Sec.~\ref{sec:ED_results} that for the case of the 18A cluster, the Chern number in parts of the CDW phase displays a surprising value of $C=1$, i.e., a spontaneous symmetry breaking occurs, in line with what has been found in related spinful models possessing a checkerboard potential instead~\cite{Vanhala16,Tupitsyn19}. To better understand this result for this \textit{finite} cluster, we describe in more detail the procedure we follow in order to compute the topological invariant. The calculation is based on the prescription presented in Refs.~\cite{Fukui05, Varney11}, where a discretized version of Eq.~\ref{eq:C} is employed. For that, we introduce twisted boundary conditions~\cite{Niu85,Didier91}, in which the many-body ground state $|\Psi^0_{\phi_x,\phi_y}\rangle$ is obtained on a torus $\{\phi_x,\phi_y\} \in [0, 2\pi)$. In the numerical computations, this range is discretized in $N_x$ and $N_y$ intervals, resulting in $\phi_x = \frac{2\pi m}{N_x}$ and $\phi_y = \frac{2\pi n}{N_y}$, with the integers $m,n$ chosen such that $m \in [0, N_x)$ and $n \in [0, N_y)$, and the ground state in such points is specified as $|\Psi^0_{m,n}\rangle$.

\begin{figure}[t]
\centering
\includegraphics[width=0.5\textwidth]{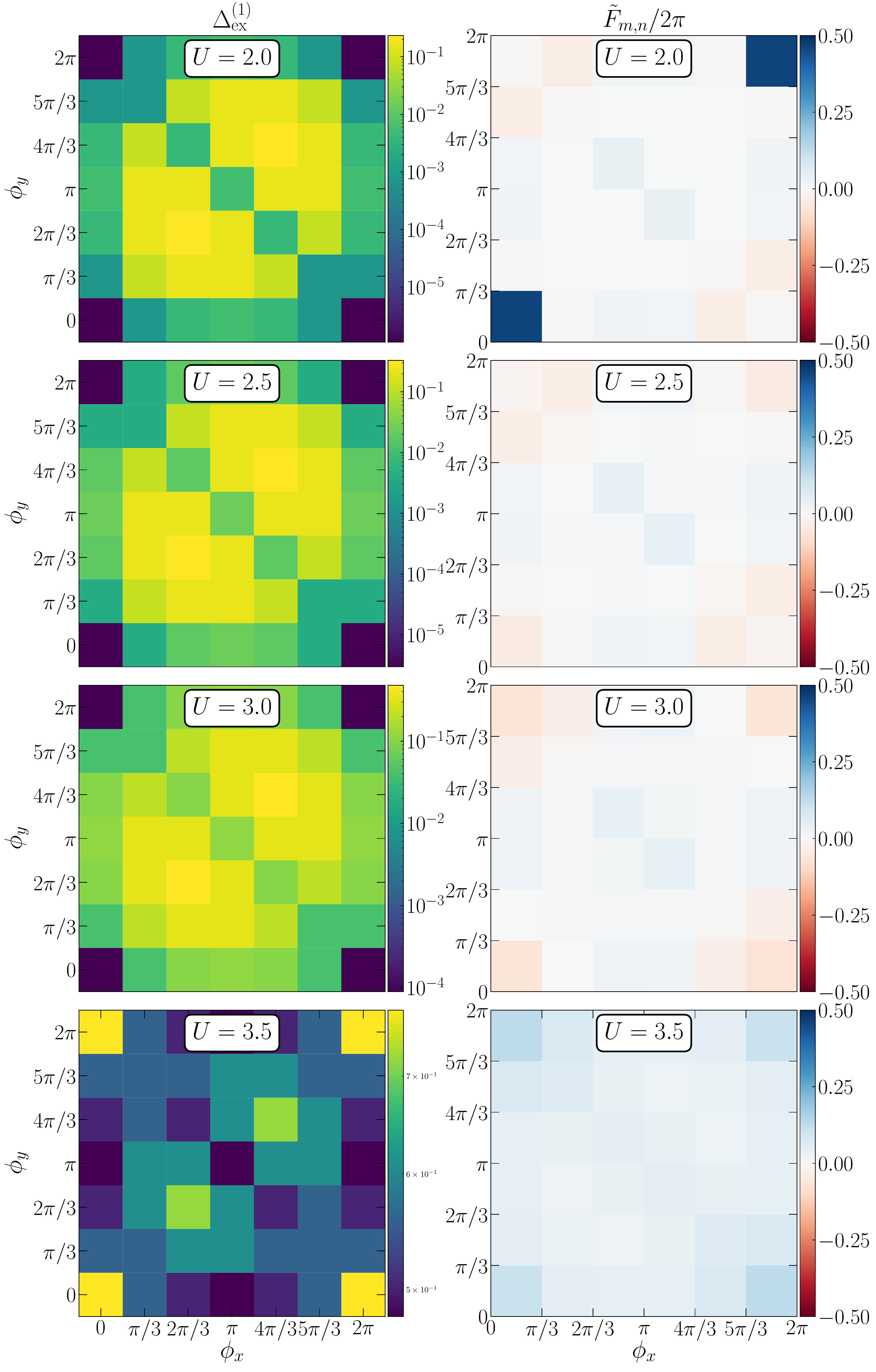}
\caption{Left panels display the excitation gap between the ground state and the first-excited many-body states $\Delta_{\rm ex}^{(1)}$ across the torus formed by the TBCs, $\{\phi_x,\phi_y\}$, for $V=1.5$. Similarly, the right panels depict the corresponding Berry curvature, normalized by $2\pi$. From top to bottom, the on-site interactions are $U= 2, 2.5, 3, 3.5$, obtained for the 18A cluster.}
\label{fig_app_1}
\end{figure}

As a result, the discrete version of the Berry curvature can be written as
\begin{equation}
 \tilde F_{m,n} = -{\rm i}\log\left(\frac{U^x_{m,n} U^y_{m+1,n}}{U^x_{m,n+1}U^y_{m,n}}\right),
\end{equation}
where the complex numbers $U_{m,n}^{x(y)}$ are the normalized overlaps of the wave functions in consecutive points of the patched torus,
\begin{align}
 U_{m,n}^x = \frac{\langle \Psi^0_{m,n}|\Psi^0_{m+1,n}\rangle}{|\langle \Psi^0_{m,n}|\Psi^0_{m+1,n}\rangle|}, && U_{m,n}^y = \frac{\langle \Psi^0_{m,n}|\Psi^0_{m,n+1}\rangle}{|\langle \Psi^0_{m,n}|\Psi^0_{m,n+1}\rangle|},
\end{align}
with $\tilde F_{m,n}$ chosen in the branch $(-\pi,\pi]$.

Finally, the topological invariant is thus written as a normalized summation of the Berry curvatures,
\begin{equation}
 C = \sum_{m,n} \frac{\tilde F_{m,n}}{2\pi},
 \label{eq:disc_C}
\end{equation}
which, for a sufficiently large discretization $N_{x,y}$, converges to the correct Chern number.

\begin{figure}[t]
\centering
\includegraphics[width=0.5\textwidth]{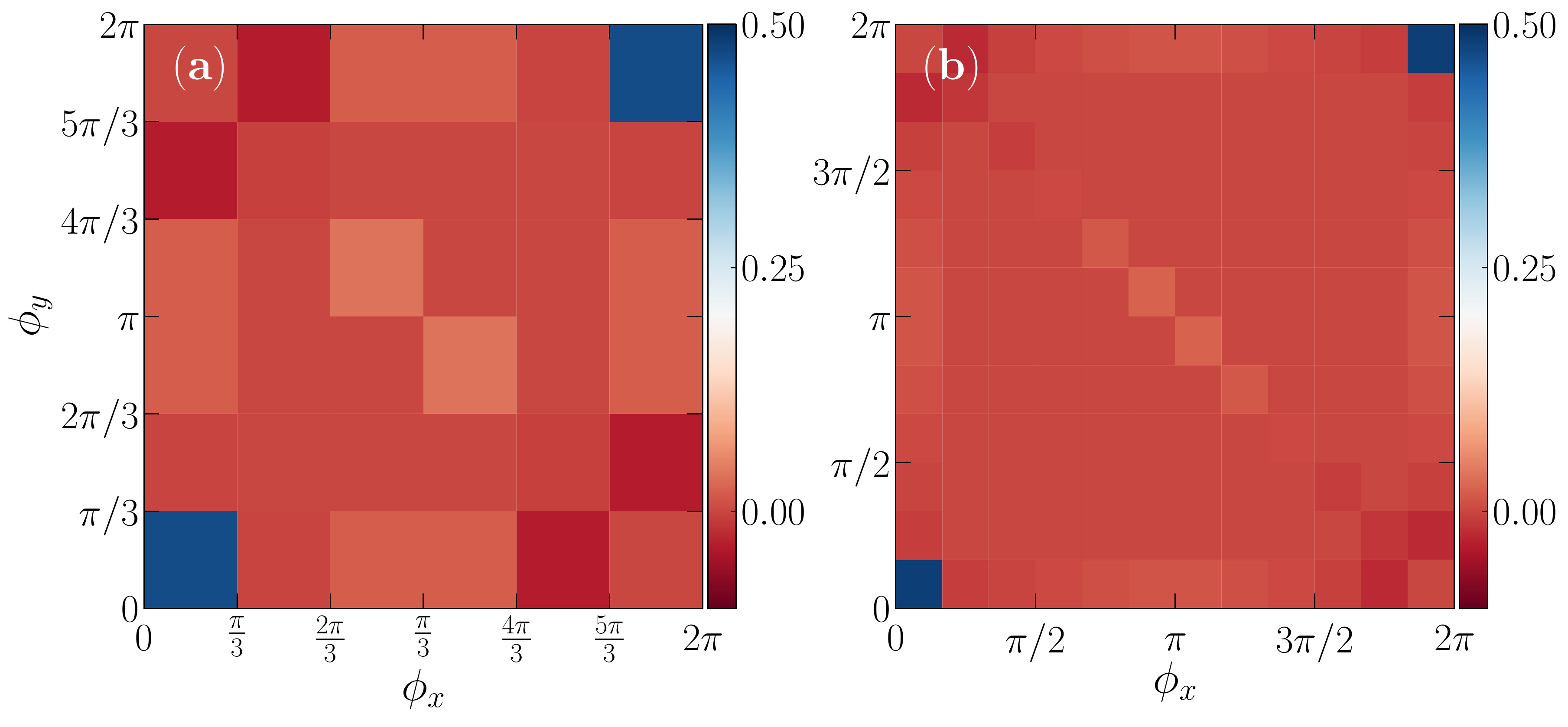}
\caption{The Berry curvature normalized by $2\pi$, $\tilde F_{m,n}/2\pi$, for the cluster 18A with parameters $U = 2$ and $V = 2$, along the TBC torus $\{\phi_x,\phi_y\}$. In (a), $N_{x,y}=6$, whereas in (b) $N_{x,y}=12$. General features are maintained albeit the much finer mesh in the latter, and the resulting Chern number is $C=1$ in both cases [see Fig.~\ref{fig_3}(a)].}
\label{fig_app_2}
\end{figure}

A necessary condition for the validity of this method is that the first-excitation gap $\Delta_{\rm ex}^{(1)}$ is always finite along the torus formed by the phases $\{\phi_x,\phi_y\}$. That is, the phases do not yield a gap closing; otherwise the Berry curvature defined above would display a singularity. In Fig.~\ref{fig_app_1}, left panels, we show an example of $\Delta_{\rm ex}^{(1)}(\phi_x,\phi_y)$ when patching $\{\phi_x,\phi_y\}$ using $N_x=N_y=6$. The four consecutive $\Delta_{\rm ex}^{(1)}$'s are chosen across a cut in the phase diagram with $V=1.5$ and on-site interactions $U = 2.0, 2.5, 3.0$ and 3.5. In this range, the computed Chern number from Eq.~\ref{eq:disc_C} is given, respectively, by $C = 1, 0, 0$ and 2. From $U=2$ to $U=2.5$, although gaps are all finite in $\{\phi_x,\phi_y\}$ (and much larger than the tolerance on the convergence set in the Krylov-Schur diagonalization), the corresponding Berry curvatures present a systematic change; see the two top right panels in Fig.~\ref{fig_app_1}. In turn, the second change of the Chern number along this $V = 1.5$ line is less surprising, and is related to the first-order phase transition, when the nearly degenerate doublet of states in the CDW phase crosses the third lowest eigenvalue in the Hamiltonian, entering in the Chern insulating phase.

We have further tested the unexpected $C=1$ to $C=0$ transition for other values of $V$, but it results in similar outcomes. Another possibility that may explain such SU(2) symmetry-broken phase on cluster 18A is related to the small number of patches $N_{x,y}$ used in the calculation of the topological invariant. Figure~\ref{fig_app_2} displays a direct comparison of the Berry curvature for a typical point in the phase diagram that resulted in $C=1$: $(U,V) = (2,2)$. Increasing the number of patches from $N_{x,y}=6$ to 12 does not alter the computed Chern number, so this technical aspect is not responsible for its appearance in this finite cluster.

%

\end{document}